\documentclass[aps,prl,preprint,superscriptaddress,showpacs]{revtex4-1}
\usepackage{amsfonts,amsmath,amssymb}
\usepackage{epsfig}
\usepackage{hyperref}
\begin{document}

\title{Temperature Control of Ion Guiding Through Insulating Capillaries}

\author{G. Kowarik}
\affiliation{Institute of Applied Physics, TU Wien - Vienna University of Technology, 1040 Vienna, Austria, EU}
\author{R. J. Bereczky}
\affiliation{Institute of Nuclear Research of the Hungarian Academy of Sciences, (ATOMKI), 4001 Debrecen, Hungary, EU}
\author{F. Ladinig}
\affiliation{Institute of Applied Physics, TU Wien - Vienna University of Technology, 1040 Vienna, Austria, EU}
\author{D. Schrempf}
\affiliation{Institute of Applied Physics, TU Wien - Vienna University of Technology, 1040 Vienna, Austria, EU}
\author{C. Lemell}
\affiliation{Institute of Theoretical Physics, TU Wien - Vienna University of Technology, 1040 Vienna, Austria, EU}
\author{J. Burgd\"orfer}
\affiliation{Institute of Theoretical Physics, TU Wien - Vienna University of Technology, 1040 Vienna, Austria, EU}
\author{K. T\H{o}k\'{e}si}
\affiliation{Institute of Nuclear Research of the Hungarian Academy of Sciences, (ATOMKI), 4001 Debrecen, Hungary, EU}
\author{F. Aumayr}
\email{aumayr@iap.tuwien.ac.at}
\affiliation{Institute of Applied Physics, TU Wien - Vienna University of Technology, 1040 Vienna, Austria, EU}

\date{\today}

\begin{abstract}
Guiding of highly charged ions (HCI) through tilted capillaries promises to develop into a tool to efficiently collimate and focus low-energy ion beams to sub-micrometer size. One control parameter to optimize guiding is the residual electrical conductivity of the insulating material. Its strong (nearly exponential) temperature dependence is the key to transmission control and can be used to suppress transmission instabilities arising from large flux fluctuations of incident ions which otherwise would lead to Coulomb blocking of the capillary.\\
We demonstrate the strong dependence of transmission of Ar$^{9+}$ ions through a single macroscopic glass capillary on temperature and ion flux. Results in the regime of dynamical equilibrium can be described by balance equations in the linear-response regime.
\end{abstract}

\maketitle

Placing or implanting a single ion at a desired point of a substrate surface with nanometer-scale precision would be highly desirable for novel applications as e.g.\ nano-modifications of surfaces \cite{rat99,akc07,said08,hel08}, fabrication of solid state qubit arrays \cite{schen02,mort08} or nano-surgery of living cells \cite{iwai08}. Slow highly charged ions (HCI) are of particular interest due to their high potential energy which is primarily deposited in a nanometer-sized volume around the impact site resulting in the emission of a large number of secondary particles \cite{aum93,andere}. Impacting on insulating materials, slow HCI may induce the formation of stable hillock- or crater type nanostructures \cite{said08,hel08,tona07,tona08}. While the emission of a large number of electrons allows detection of each ion impact with unit efficiency and therefore single ion hit monitoring \cite{rsi}, the morphology and size of the resulting material modification can be tuned by the charge state of the incoming highly charged ion \cite{aum08}. The main challenge remaining is to define the ion impact point as precisely as possible. One possibility for the preparation of a well-focused HCI nano-beam lies in the utilization of the so-called capillary-guiding effect \cite{stol02,saha06,stol07,hell07,skog08,stol08,stol09,kan09,ber09,nak09,stol10,zhan10} using tapered capillaries with sub-micrometer exit diameters \cite{iwai08,nak09,ike06,cas08}.

First experiments on guiding HCI through straight insulator nano-capillaries showed a remarkable effect: after an initial charge up phase, the ion beam could be steered by tilting the capillary axis while remaining in the initial charge state indicating that the transmitted ions never touched the inner walls. Subsequent experiments confirmed this guiding effect also for macroscopic glass capillaries, both straight and tapered ones, suggesting tapered glass capillaries as funnels for HCI beams with unprecedented guiding and focusing properties.

Microscopic simulations for nanocapillaries \cite{schie05_2,schie07,lem07,schie09} revealed that a self-organized charge up of the capillary walls due to preceding HCI impacts leads to an electric guiding field, which steers the incoming projectile ions along the capillary axes. Ion guiding ensues as soon as a dynamical equilibrium of charge-up by the ion beam and charge relaxation by bulk or surface conductivity is established. These simulations showed that a stable transmission regime required a delicate balance between incident ion flux and charge relaxation via surface and bulk conduction, conditions, which were obviously met in almost all cases studied experimentally so far. While the conceptual understanding of the processes leading to ion guiding has progressed well, application of this technique as a tool for ion beam formation has emerged only recently and the search for tuning parameters to control and optimize HCI transmission has begun.

In this letter we show that a key control parameter for guiding is the small residual electric conductivity of the highly insulating capillary material whose dependence of temperature $\sigma(T)$ is nearly exponential \cite{love83}. Therefore, guiding can be tuned by moderate temperature variation near room temperature as proposed in \cite{bun08}. We demonstrate that increasing the temperature of a glass capillary and therefore its conductivity leads to a reduction of guiding and, eventually, to a complete disappearance of the guiding effect. The strong temperature dependence can be employed to stabilize guiding against Coulomb blocking due to a high incident ion flux \cite{nak09,kre11}. 

We use a single straight macroscopic glass capillary (inner diameter: 160 $\mu$m; length: 11.4 mm) made of Borosilicate (Duran\texttrademark) for which the guiding effect has been previously established \cite{ber09}. The current experimental set-up allows for a controlled and uniform temperature variation of the glass capillary. An oven made of massive copper parts surrounds the capillary in order to guarantee a uniform temperature distribution along the entire tube. The temperature of the copper parts is monitored by a K-Type thermocouple. Stainless steel coaxial heaters surrounding the oven are used for heating. Temperatures below room temperature are achieved by cooling the sample holder via heat-conduction to a massive UHV copper feed-through connected to a liquid nitrogen bath outside the UHV chamber, while the sample temperature is stabilized by the PID controlled heaters. A temperature range from $-30^\circ$C up to $90^\circ$C ($243\leq T\leq 363$ K) can be probed. Within such a moderate variation $\Delta T/T\approx 0.3$ the conductivity varies by almost five orders of magnitude. All measurements are performed under UHV conditions at a base pressure below $5 \times 10^{-9}$ mbar. Ar$^{7+}$ and Ar$^{9+}$ ions with a kinetic energy of 4.5 keV are provided by the ECR ion source in Vienna \cite{gal07}. The extracted ions are focused, mass-to-charge-separated, collimated to a divergence angle of less than $\pm 0.5^\circ$, and eventually hit a metallic entrance aperture directly in front of the capillary. This entrance aperture has a diameter of about 120 $\mu$m. The beam-spot diameter at this position is about 2.5 mm. For beam diagnostic and monitoring purposes, a reference aperture (100 $\mu$m diameter) can be inserted into the beam instead of the capillary. Transmitted ions hit a 50 mm diameter position sensitive micro-channel-plate detector with wedge-and-strip anode, located about 18 cm behind the sample. Charge state analysis of the transmitted ions is possible be means of a pair of electrostatic deflector plates located near the exit of the capillary. Transmission rates are recorded after steady-state conditions (i.e.\ a dynamical equilibrium) are reached. We emphasize that the existence of steady-state guiding can only be reached within a limited parameter range.

\begin{figure}[ht]
\centerline{\epsfig{file=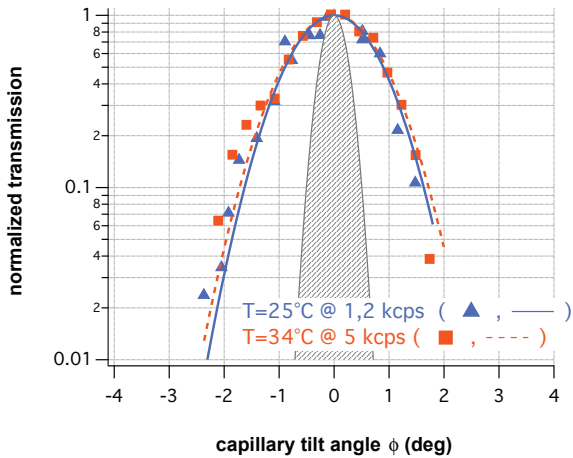,width=8cm}}
\caption{(Color online): Comparison of normalized transmission curves of 4.5 keV Ar$^{9+}$ ions guided through a glass capillary differing by a factor of $\sim 4$ in the projectile flux and the conductivity (due to a temperature difference of $9^\circ$ C) resulting in the same control parameter $\alpha$. Gaussian fits through the data (solid line: room temperature; dashed line: $34^\circ$ C) show an almost identical transmission profile. The shaded area indicates the angular distribution of the transmitted beam through the geometric opening in the absence of guiding.}
\label{fig1}
\end{figure}

Transmission as a function of tilt angle $\phi$ is measured starting at $\phi=0^\circ$ as determined by geometrically maximizing the transmitted intensity. The tilt angle is then stepwise increased with $\Delta\phi < 1^\circ$ until transmission becomes negligible. Subsequently, starting at this deflection angle, the capillary is tilted back stepwise and eventually into the opposite direction (negative $\phi$). For each tilt angle the total ion count rate onto the detector is summed up and a dead-time correction is applied. Finally, the transmission curves are normalized with respect to the transmission at $\phi=0^\circ$ (Fig.\ \ref{fig1}).

The present analysis of the transmission proceeds within the generalization of a rate-equation model which has been previously tested by comparison with a microscopic transport simulation for nanocapillaries\cite{schie05_2}. We note that a corresponding microscopic simulation for transmission through macrocapillaries is out of reach. The balance equation for the deposited charge $Q(t)$ at the capillary wall controlling guiding reads
\begin{equation}
\frac{dQ(t)}{dt}=\left( j_\text{in}-j_\text{tr}-\frac{Q(t)}{\tau_\text{eff}}\right) + \left(\frac{dQ}{dt}\right)_\text{stoc}
\label{eq1}
\end{equation}
where $j_\text{in}$ is the incoming ion current, $j_\text{tr}$ is the current transmitted through the capillary, and $\tau_\text{eff}$ is the effective discharging time related to the linear-response surface and bulk conductivities
\begin{equation}
\tau_\text{eff}^{-1}(T)\approx c\cdot\sigma_\text{surf}(T)+c\cdot\sigma_\text{bulk}=c\cdot\sigma_\text{eff}(T)\label{eq2}
\end{equation}
which is strongly temperature ($T$) dependent but independent of the charge $Q$ present. For later reference we have added a stochastic non-linear discharge term [$(dQ/dt)_\text{stoc}$ in Eq.\ \ref{eq1}] which will be, in general, a functional of the charging history, $Q(t')\ldots t'\leq t$, and contributes above a critical value for charging $Q(t)>Q_\text{crit}$. When a unique stable dynamical equilibrium can be reached this term can be neglected. However, in cases where multivalued hysteresis-like transmission properties appear \cite{ike07} or no steady-state can be reached, such corrections become important.

Expressing $j_\text{tr}$ in terms of the tilt-angle ($\phi$) dependent transmission probability [$P(\phi,Q)$]
\begin{equation}
j_\text{tr}=j_\text{in}\cdot P(\phi,Q)
\end{equation}
the dynamical equilibrium charge-up $Q_\text{eq}$ follows from Eq.\ \ref{eq1} as
\begin{equation}
Q_{eq}=\frac{j_\text{in}}{c\sigma_\text{eff}(T)}\{1-P(\phi,Q_\text{eq})\}\, .
\end{equation}
The parameter controlling guiding under equilibrium condition, $\alpha(T)=j_\text{in}/c\sigma_\text{eff}(T)$ is thus the ratio of incoming current to conductivity. Consequently, transmission properties should remain unchanged for simultaneously varying $j_\text{in}$ and $\sigma_\text{eff}(T)$ provided that $\alpha(T)$ is kept fixed. The normalized transmission of HCI through the glass capillary (Fig.\ \ref{fig1}) displays, indeed, an almost identical angular dependence fitted to a Gaussian, $P(\phi,Q_\text{eq})\propto\exp (-\phi^2/\phi_g^2)$ with $\phi_g$ a characteristic guiding angle of $\phi_g=1.25^\circ$ when varying both flux and conductivity by a factor 4. Note that $\phi_g$ exceeds the geometric opening angle $\phi_0$ by a factor 4 clearly indicating that charge-up induced guiding takes place.

\begin{figure}[ht]
\centerline{\epsfig{file=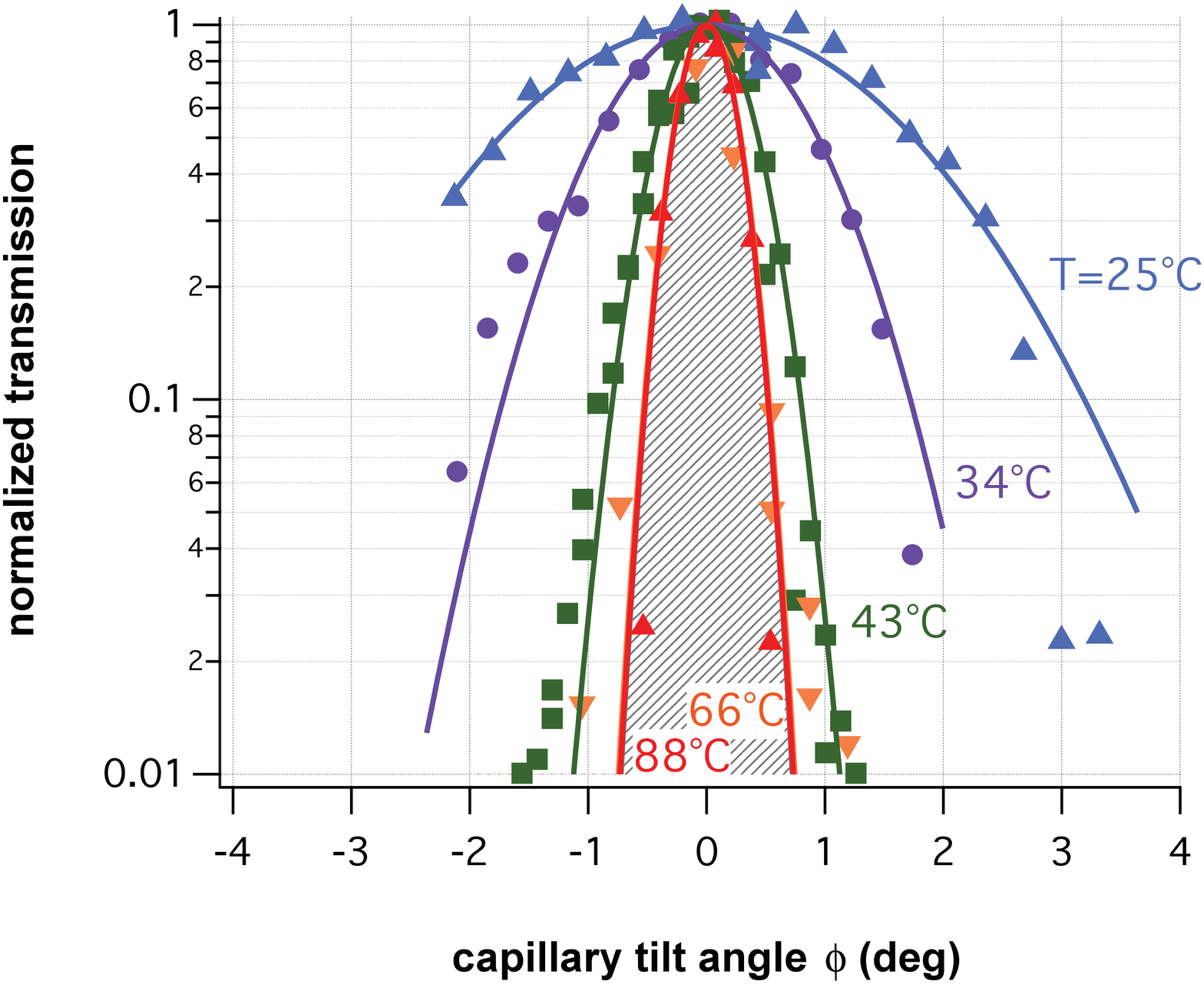,width=8.5cm}\epsfig{file=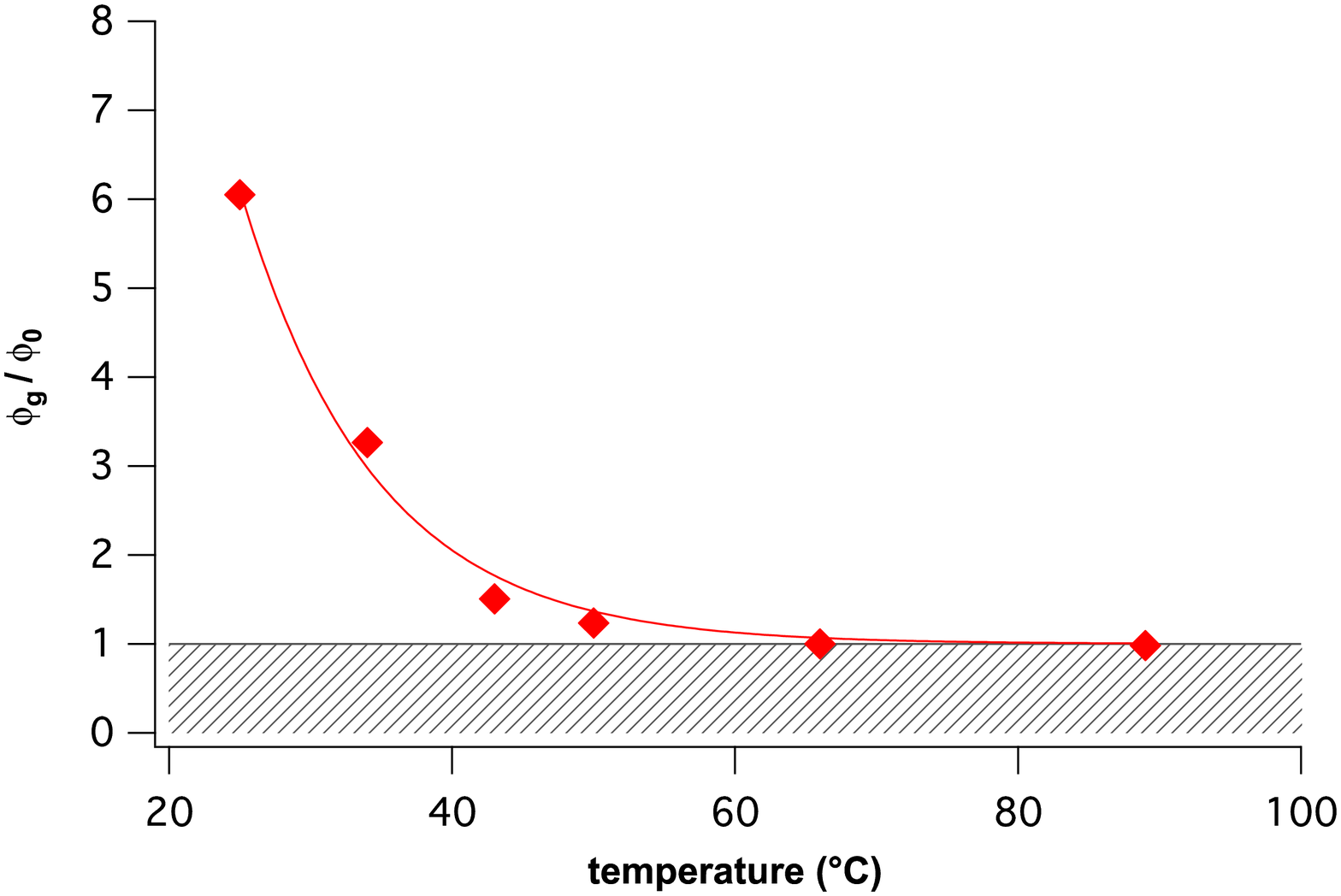,width=6.5cm}}
\caption{(Color online): left: Normalized transmission curves for 4.5 keV Ar$^{9+}$ ions guided through a glass capillary for different temperatures ranging from 24 to $88^\circ$ C. The flux of the incident 4.5 keV Ar$^{9+}$ ions was kept constant at about 5000 ions entering the capillary per second. Gaussian fits through the data points are shown as solid lines. The shaded area indicates the geometric limit of transmission in the absence of guiding. right: Guiding angle $\phi_g$ over geometric opening angle $\phi_0$ as a function of temperature (data derived from Fig.\ \ref{fig2}a).}
\label{fig2}
\end{figure}

Keeping the incident current constant but increasing the temperature will lead to a lower equilibrium value for the total charge $Q_\text{eq}$ due to the increased surface and bulk conductivities. As a consequence, $j_\text{tr}$ or, equivalently, the guiding angle $\phi_g$ will be reduced (Fig.\ \ref{fig2}). 
The angular distribution of the normalized transmission current varies drastically with temperature. The temperature range of $24-88^\circ$ C ($297\leq T\leq 361$ K) covers almost 3 orders of magnitude in conductivity \cite{love83}. At these high temperatures ($T\geq 340$ K) the removal of deposited charges by the elevated conductivity is so efficient and $Q_\text{eq}$ (Eq.\ \ref{eq2}) so small that $P(\phi,Q_\text{eq})$ approaches zero outside the geometric opening angle or, equivalently, $\phi_g\to\phi_0$. By increasing the temperature, guiding can be controlled and \textit{reversibly} switched off.

Increasing the conductivity by increasing the temperature opens a pathway to overcome one of the limitations current experimental charged-particle transmission studies face: state-of-the-art cleaning methodology of surface science such as sputtering with keV rare-gas ions cannot be applied due to immediate charge-up of the inner surface. Collisional removal of surface deposits (such as hydrocarbons, water, or chemical residues from etching) is suppressed due to deflection at the Coulomb mirror at large distances or Coulomb blocking of the capillary. The present result shows that for temperature-resistant materials such as glass, SiO$_2$, or Al$_2$O$_3$ sputtering at temperatures $T\geq T_0$ where the dynamical equilibrium for charge-up at incoming currents $j_\text{in}$ sufficient for sputtering lies below the threshold for guiding, removal of surface coverage by sputtering becomes possible.

\begin{figure}[h]
\centerline{\epsfig{file=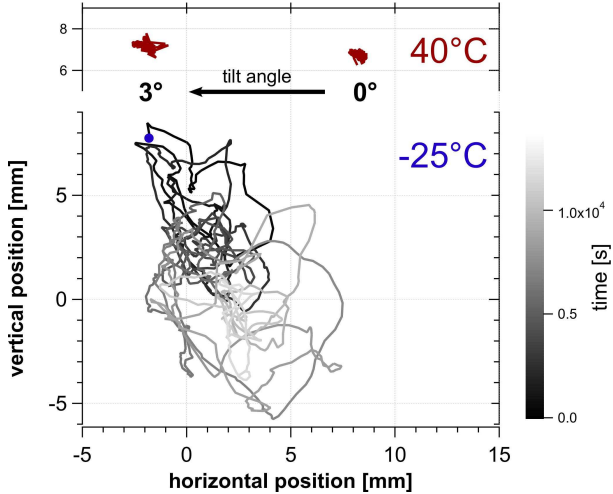,width=8cm}}
\caption{(Color online): Position of the beam spot on the detector after passage through the capillary. Upper part: stable beam spot under guiding conditions ($T=313$ K) for tilt angle $\phi=3^\circ$ (and $\phi=0^\circ$ for reference); lower part: random walk of beam spot under non-equilibrium conditions ($T=248$ K) for tilt angle $\phi=3^\circ$. The beam position has been evaluated by a 2D Gaussian fit of the time-binned spatial distribution on the detector. 4.5 keV Ar$^{7+}$ ions have been used as projectiles.}
\label{fig3}
\end{figure}

In the opposite limit for large $\alpha$, i.e.\ either by further reducing the conductivity or increasing the current, the charge deposition increases beyond $Q_\text{crit}$ reaching the regime of non-linear response. One indication is that a unique dynamical equilibrium can no longer be established. For $Q(t)>Q_\text{crit}$, the stochastic discharge term in Eq.\ \ref{eq1} significantly contributes. The onset of dynamical instability can be directly monitored by the stochastic motion of the beam spot (Fig.\ \ref{fig3}).
While in the linear stable regime the beam spot remains stationary and well-defined at the position corresponding to the guiding angle, at low temperatures the beam spot undergoes a random motion that does not converge towards a stationary beam spot. We interpret that in terms of the transient formation of charge patches fluctuating in charge amplitude and position (most likely close to the exit). We note that such non-equilibrium behavior more likely to be observed in the experiments on individual macrocapillaries such as the present one or for funnel geometries \cite{nak09}. For experiments employing ensembles of the order of $10^6$ nanocapillaries, individual capillaries may also undergo stochastic fluctuations of Coulomb blocking or deflection angle. On the ensemble average, nevertheless, a beam spot stable in magnitude and position may result. For macrocapillaries, on the other hand, complete Coulomb blocking by charge-up is unlikely due to the large total charge required. Instead, randomly fluctuating patches resulting from charging, discharging, and charge migration is the more likely scenario for the non-equilibrium state of transmission.

Summarizing, we have presented first results on the temperature dependence of macrocapillary transmission which opens the pathway to improved control of ion-beam guiding using nano- and microcapillaries. Exploiting the strong temperature dependence of the electrical surface and bulk conductivities (about 1 order of magnitude per $\Delta T=25$ K temperature change) allows for optimization of the transmitted beam either for transmitted current and guiding angle of the transmitted beam. Transmission properties of insulating capillaries are governed by the ratio of incident current (charging) and conductivity (discharging) $\alpha=j_\text{in}/c\sigma_\text{eff}(T)$. The improved control over the dynamical equilibrium of charges on the inner capillary wall required for stable transmission conditions due to the strong temperature dependence of $\sigma_\text{eff}(T)$ opens up the possibility of improved surface preparation of inner capillary walls. Conventional sputter guns may be used to clean the surface while keeping the capillary surface uncharged at the same time. This holds the promise for detailed investigations of the influence of the capillary material on ion transmission.

\begin{acknowledgments}
This work has been supported by the European Project ITS-LEIF (No. RII3\#026015), by  Austrian Science Foundation FWF, and in part by TeT under Grant No. AT-2/2009.
\end{acknowledgments}

\end{document}